\documentclass[10pt,journal,compsoc,onecolumn]{IEEEtran}

\usepackage{chemformula}
\usepackage{float}
\usepackage{algorithm}
\usepackage{algpseudocode}
\usepackage{amsmath}
\usepackage[utf8]{inputenc} 
\usepackage[T1]{fontenc}    
\usepackage{hyperref}       
\usepackage{url}            
\usepackage{booktabs}       
\usepackage{amsfonts}       
\usepackage{nicefrac}       
\usepackage{microtype}      
\usepackage{siunitx}
\DeclareSIUnit{\calorie}{cal}
\DeclareSIUnit{\kilocalorie}{kcal}  %
\usepackage{fancyhdr}       
\usepackage{graphicx}       
\graphicspath{{imgs/}}     
\usepackage{lineno}
\begin{document}
\title{CreoPep: A Universal Deep Learning Framework for Target-Specific Peptide Design and Optimization}

\author{

\IEEEauthorblockN{
\raggedright
    Cheng Ge\textsuperscript{1,2},
    Han-Shen Tae\textsuperscript{3},
    Zhenqiang Zhang\textsuperscript{1,2},
    Lu Lu\textsuperscript{1,2},
    Zhijie Huang\textsuperscript{1,2},
    Yilin Wang\textsuperscript{4},
    Tao Jiang\textsuperscript{1,2},
    Wenqing Cai\textsuperscript{5},
    Shan Chang\textsuperscript{6},
    David J. Adams\textsuperscript{3},
    Rilei Yu\textsuperscript{1,2,\IEEEauthorrefmark{1}}
}

\IEEEauthorblockA{
    \raggedright
    \textsuperscript{1}Key Laboratory of Marine Drugs, Chinese Ministry of Education,
    School of Medicine and Pharmacy, Ocean University of China, Qingdao 266003, China\\
    \textsuperscript{2}Laboratory for Marine Drugs and Bioproducts,
    Qingdao Marine Science and Technology Center, Qingdao 266237, China\\
    \textsuperscript{3}Molecular Horizons, Faculty of Science, Medicine and Health,
    University of Wollongong, Wollongong, NSW 2522 Australia\\
    \textsuperscript{4}Faculty of Information Science and Engineering,
    Ocean University of China, Qingdao 266100, China\\
    \textsuperscript{5}Shandong Academy of Pharmaceutical Sciences, Jinan, China\\
    \textsuperscript{6}Institute of Bioinformatics and Medical Engineering, School of Electrical and Information Engineering, Jiangsu University of Technology, Changzhou 213001, China
}

    \thanks{\IEEEauthorrefmark{1}Corresponding author: Rilei Yu. E-mail: ryu@ouc.edu.cn}
}

\IEEEtitleabstractindextext{%
\begin{abstract}
Target-specific peptides, such as conotoxins, exhibit exceptional binding affinity and selectivity toward ion channels and receptors. However, their therapeutic potential remains underutilized due to the limited diversity of natural variants and the labor-intensive nature of traditional optimization strategies. Here, we present CreoPep, a deep learning-based conditional generative framework that integrates masked language modeling with a progressive masking scheme to design high-affinity peptide mutants while uncovering novel structural motifs. CreoPep employs an integrative augmentation pipeline, combining FoldX-based energy screening with temperature-controlled multinomial sampling, to generate structurally and functionally diverse peptides that retain key pharmacological properties. We validate this approach by designing conotoxin inhibitors targeting the $\alpha$7 nicotinic acetylcholine receptor, achieving submicromolar potency in electrophysiological assays. Structural analysis reveals that CreoPep-generated variants engage in both conserved and novel binding modes, including disulfide-deficient forms, thus expanding beyond conventional design paradigms. Overall, CreoPep offers a robust and generalizable platform that bridges computational peptide design with experimental validation, accelerating the discovery of next-generation peptide therapeutics.
\end{abstract}

}

\maketitle

\IEEEdisplaynontitleabstractindextext

%
\IEEEpeerreviewmaketitle

\section*{Introduction}
\begin{figure}[h]
\centering
\includegraphics[width=0.9\textwidth]{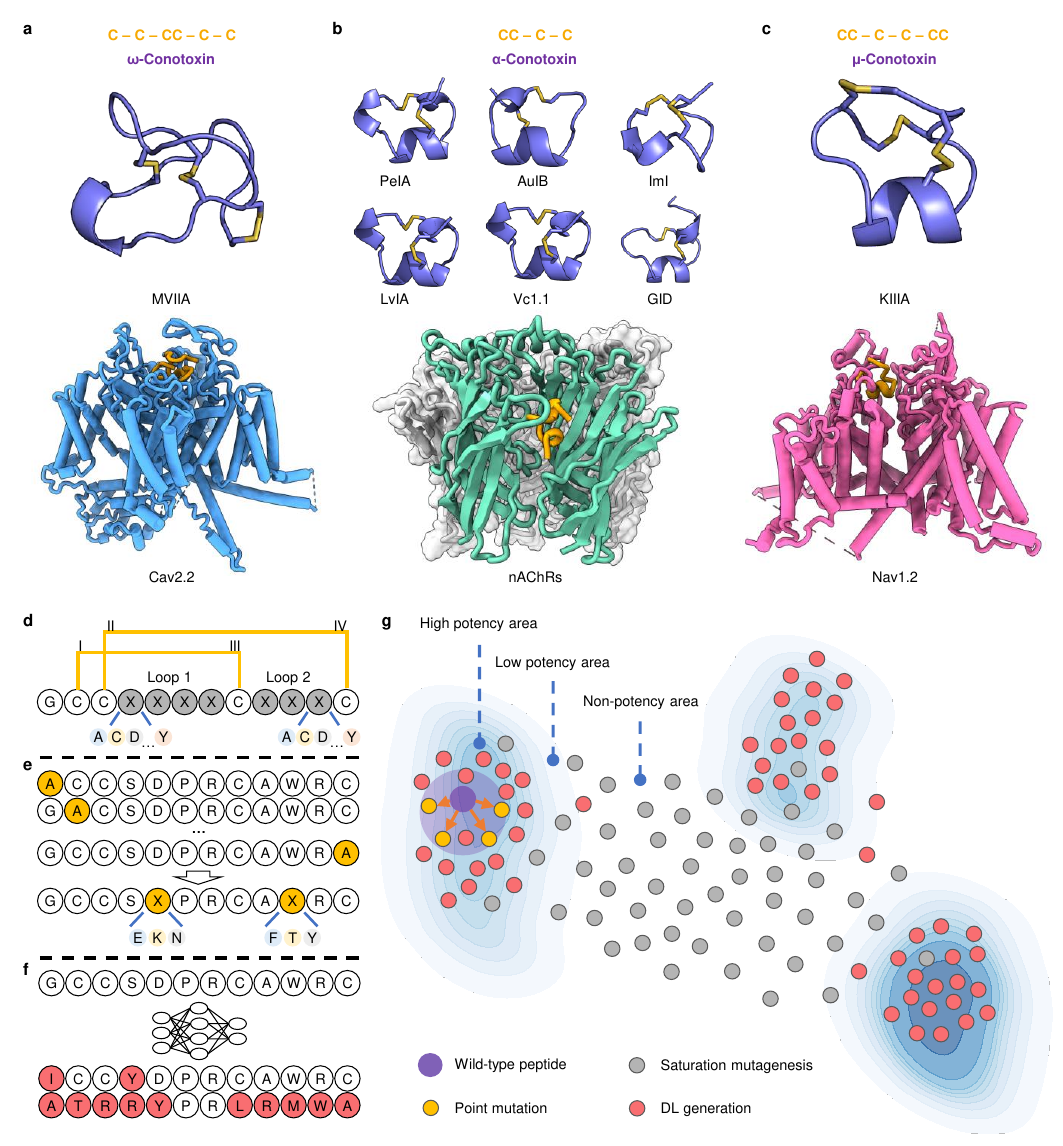}
\caption{\textbf{Overview of mutation strategies for target-specific peptides.} \textbf{a}, $\omega$-Conotoxin features three disulfide bonds. An example is $\omega$-conotoxin MVIIA, which targets Cav2.2 channels (PDBid: 7MIX). \textbf{b}, $\alpha$-Conotoxin contains two disulfide bonds. Examples include $\alpha$-conotoxin ImI and Vc1.1, which primarily target nAChRs (PDBid: 7KOO). \textbf{c}, $\mu$-Conotoxin has three disulfide bonds. An example is $\mu$-conotoxin KIIIA, which blocks voltage-gated sodium channels (PDBid: 6J8E). \textbf{d}, Saturation mutagenesis, where each amino acid residue in the loop1 and loop2 regions of the ImI sequence is replaced by one of the other 19 natural amino acids. Yellow lines indicate disulfide bonds. \textbf{e}, Point mutations are introduced at specific positions identified through alanine scanning, targeting either key or non-key or key residues. \textbf{f}, Target-specific peptide design based on deep learning, directly generating high-potency ImI mutants. \textbf{g}, Schematic diagram of sequence space representation of ImI mutants. The purple dot represents the wild-type peptide, the orange dots represent peptides generated by point mutations, the large purple circle indicates the exploration range of point mutations, the gray dots represent peptides produced by saturation mutagenesis, and the pink dots represent peptides generated by deep learning (DL) methods.
}\label{fig1}
\end{figure}

Target-specific peptides are a class of biomolecules that bind with high affinity and specificity to biological targets such as receptors or ion channels, modulating their function (Fig.~\ref{fig1}a-c) \cite{ref1,ref2}. Among them, conotoxins, bioactive peptides derived from marine cone snail venom, exhibit remarkable potency and selectivity across various ion channel families, including voltage-gated calcium (Cav) channels, nicotinic acetylcholine receptors (nAChRs), and voltage-gated sodium (Nav) channels \cite{ref3,ref4,ref5,ref6}. Their precision in regulating neural activity makes conotoxins promising scaffolds for drug development \cite{ref3,ref7}.

Conotoxins are stabilized by conserved disulfide bonds and organized into at least 30 distinct structural frameworks, each associated with specific pharmacological targets \cite{ref8}. For example, $\omega$-conotoxins target Cav channels \cite{ref9} (Fig.~\ref{fig1}a), $\alpha$-conotoxins act on nAChRs \cite{ref10} (Fig.~\ref{fig1}b), and $\mu$-conotoxins modulate Nav channels \cite{ref11} (Fig.~\ref{fig1}c). Despite their conserved structural scaffolds, conotoxins exhibit extraordinary sequence diversity, with an estimated 1 million bioactive variants existing in nature. However, fewer than 1\% ($\sim$10,000) have been sequenced, and only a small subset pharmacologically characterized. This vast diversity highlights their therapeutic potential \cite{ref3,ref12,ref13}, as demonstrated by $\omega$-conotoxin MVIIA  \cite{ref14,ref15}, a 25-residue peptide from \textit{Conus magus} venom that selectively inhibits N-type calcium channels and has been approved for chronic pain treatment. However, naturally derived peptides often suffer from suboptimal activity, off-target effects, and poor metabolic stability, presenting significant barriers to clinical translation \cite{ref1}. Overcoming these limitations typically requires extensive mutagenesis screening and chemical modifications, a labor-intensive process that underscores the need for more efficient peptide engineering strategies \cite{ref16,ref17,ref18}.

Current strategies for peptide mutation screening fall into two main categories: saturation mutagenesis (Fig.~\ref{fig1}d) and targeted point mutation (Fig.~\ref{fig1}e). In saturation mutagenesis, specific amino acid positions are substituted with all 20 natural amino acids, followed by structural prediction (e.g., RosettaFold \cite{ref19}, AlphaFold3 \cite{ref20}) and binding affinity estimation (e.g., FoldX \cite{ref21}, FlexPepDock \cite{ref22}, HPEPDOCK \cite{ref23}). Although this approach can identify high-affinity mutants and explore broad sequence diversity, it is computationally expensive and yields relatively few viable candidates. In contrast, targeted point mutation uses techniques like alanine scanning to identify critical residues through binding assays or \textit{in vitro} activity measurements. While more efficient, this method is limited by its reliance on prior structural knowledge and its inability to broadly explore sequence diversity constraining its utility for \textit{de novo} design. Importantly, both strategies are restricted by dependence on fixed backbone conformations, limiting exploration of novel structural motifs---a key limitation in advancing peptide therapeutics.

The emergence of deep learning is beginning to transform traditional drug discovery, offering promising avenues to accelerate peptide and protein design \cite{ref24,ref25,ref26,ref27,ref28,ref29,ref30,ref31,ref32,ref33,ref34}. By training models to predict protein-peptide interactions (PpIs), researchers can rapidly screen large peptide libraries. For instance, Lei \textit{et al.} (2021) developed a Convolutional Neural Networks (CNNs) integrated with self-attention to predict PpIs from sequence data alone, while also identifying peptide binding residues \cite{ref35}. However, the utility of such models is constrained by the limited availability of high quality training data. Beyond screening, deep generative models have proven powerful in directly designing peptide analogues with enhanced affinity. Deng \textit{et al.} (2024) introduced a reinforcement learning-based model, RLpMIEC, which integrates peptide-MHC interaction energy spectra and sequence features to generate peptides with high binding affinity for MHC-I molecules \cite{ref36}. Similarly, Chen \textit{et al}. (2024) used a Gated Recurrent Unit (GRU)-based Variational Autoencoder (VAE) combined with Metropolis-Hastings (MH) sampling to design peptide inhibitors targeting $\beta$-catenin and NF-$\kappa$B essential modulator \cite{ref37}. Despite their success, these methods often rely on their respective targets, their generalizability remains constrained by dependence on target-specific training datasets and explore limited conformational diversity restricting their generalizability.

In this study, we present a generalizable approach to overcome these limitations by developing a target-specific peptide design algorithm capable of exploring active conformational space while generating high affinity mutants (Fig.~\ref{fig1}f-g). We introduce CreoPep, a conditional generative framework based on a masked language model inspired by diffusion models \cite{ref38}. CreoPep uses a Progressive Masking (PM) strategy, gradually increasing masked tokens during training (adding noise) and incrementally predicting them during generation (denoising). This enables the model to more effectively learn the relationship between peptide sequence, structure, and function. Additionally, we integrate CreoPep with a novel peptide augmentation pipeline, incorporating multiple rounds of FoldX-based binding energy screening to generate a large pseudo-labeled dataset that enhances model training. To ensure both diversity and specificity, CreoPep employs temperature-controlled multinomial sampling, enabling generation of functionally diverse yet high-affinity peptides. We validate the efficacy of CreoPep by designing peptide inhibitors targeting the $\alpha$7 nAChR, achieving encouraging results in electrophysiological assays. Our work provides a robust and generalizable framework that bridges computational peptide design and experimental validation. By harnessing deep learning and advanced generative modeling, CreoPep accelerates the discovery of target-specific peptides and opens new frontiers in therapeutic development, with broad implications for drug discovery, synthetic biology, and personalized medicine.

\section*{Results}
\subsection*{Improving conotoxin binding through mutant design}
Current conotoxin optimization strategies focus primarily on enhancing the potency of known sequences through iterative mutagenesis. For example, $\alpha$-conotoxin ImI, a compact \SI{12}{}-residue peptide, has been subjected to extensive point mutation studies, albeit with limited success, likely due to its evolutionarily constrained structure. To explore alternative optimization strategies, we first implemented saturation mutagenesis on the ImI/$\alpha$7 nAChR system. By systematically replacing residues in both loop 1 and loop 2 with all 20 natural amino acids, we generated 1,000 variants and assessed their binding free energy changes ($\Delta\Delta G$) using FoldX (Fig.~\ref{fig1}d). This screen identified 65 mutants (\SI{6.5}{\%} hit rate) with improved binding affinity relative to wild-type ImI (Supplementary File 1), demonstrating the potential but limited efficiency of this approach.

To improve the success rate, we next applied a constrained mutation strategy by fixing three experimentally validated critical residues (D5, P6, R7) while randomizing the remaining positions \cite{ref39}. This approach increased the number of favorable mutants to 253 (Supplementary File 1), confirming that incorporating structural knowledge can guide more effective optimization. However, this improvement in efficiency came at the cost of reduced exploration across the broader sequence-structure landscape. These findings highlight a fundamental trade-off in conotoxin engineering: while structural constraints enhance hit rates, they concurrently restrict access to novel structural motifs. To overcome this limitation, we aim to develop an advanced peptide design framework that balances hit rate and conformational diversity, enabling the efficient generation of functional variants while expanding the accessible chemical and structural space.

\subsection*{Development of the generative model CreoPep for conotoxin mutation design}
\begin{figure}[h]
\centering
\includegraphics[width=0.9\textwidth]{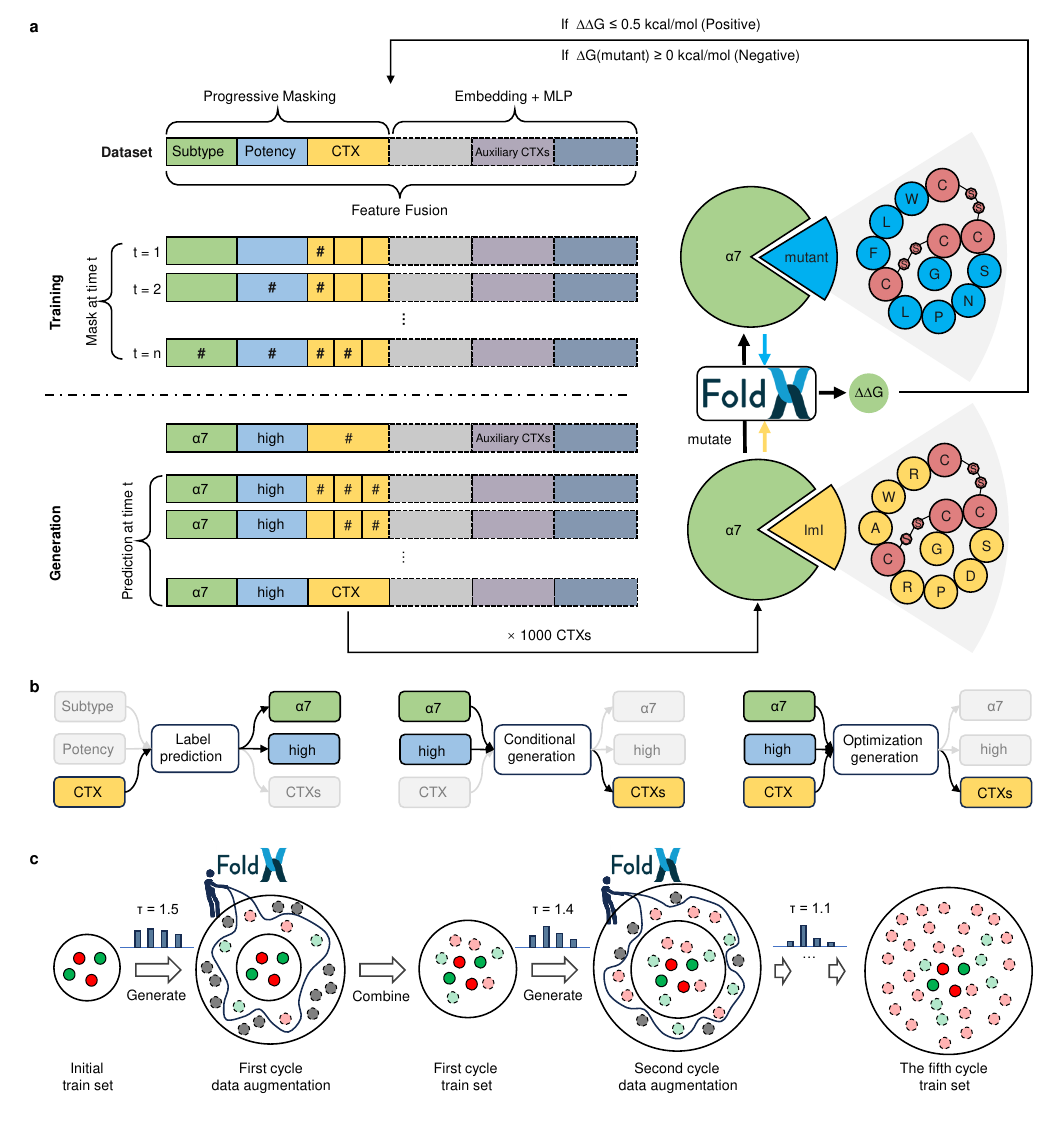}
\caption{\textbf{Overview of CreoPep.} \textbf{a}, Schematic representation of the CreoPep framework. Left panel (top to bottom): Training dataset format and feature extraction methods for each component, training phase, and generation phase. Right panel: Mutant screening workflow using FoldX for peptides generated by CreoPep. \textbf{b}, Illustration of three core functional capabilities of CreoPep. \textbf{c}, Data augmentation pipeline used to enhance training diversity and model performance. 
}\label{fig2}
\end{figure}

To overcome limitations in current conotoxin design, we developed CreoPep, a deep learning-based conditional generative model that leverages evolutionary information to efficiently explore conotoxin chemical space (Fig.~\ref{fig2}a). CreoPep is built on ProtBert’s masked language modeling (MLM) framework \cite{ref40}, pretrained on large-scale protein sequence datasets via self-supervised learning. It generates structurally diverse and pharmacologically relevant conotoxin variants through three key innovations. First, we introduced a Progressive Masking (PM) scheme that replaces the conventional fixed \SI{15}{\%} masking rate with a dynamic schedule that gradually increases the number of masked tokens during training. This enables more nuanced and context-aware learning of sequence discrepancies. During sequence generation, CreoPep uses this framework to iteratively predict one masked residue at a time until a complete conotoxin sequence is reconstructed. Second, the training data for each instance included: (1) a subtype label (one of the 53 subtypes), (2) a potency label (high or low), (3) a wild-type conotoxin, and (4) auxiliary conotoxins (randomly selected peptides share the same subtype and potency as the wild-type). Subtype and potency labels provide explicit functional constraints, while auxiliary sequences offer implicit functional patterns to guide model learning. Third, all inputs except the auxiliary conotoxins were encoded using the PM scheme; auxiliary conotoxins were instead encoded via a Multilayer Perceptron (MLP). The combined input features were fused using a Convolutional Neural Network (CNN) to capture higher-order representations.

CreoPep integrates three key tasks into a single framework: label prediction, conditional generation, and optimization generation (Fig.~\ref{fig2}b). (1) Label prediction, where subtype and potency are masked and inferred from the conotoxin sequence and optional auxiliary peptides. (2) Conditional generation, where masked residues are predicted iteratively under the guidance of target subtype/potency labels and auxiliary peptides to construct functionally tailored sequences. (3) Optimization generation, where multiple candidates are generated conditionally, evaluated via the label prediction module, and filtered to retain only those with improved potency scores and enhanced receptor-specific confidence relative to the inputs. This unified architecture enables efficient, functionally guided navigation of conotoxin sequence space, facilitating the design of novel peptide variants with improved binding affinity and functional specificity.

\begin{figure}[t!]
\centering
\includegraphics[width=0.9\textwidth]{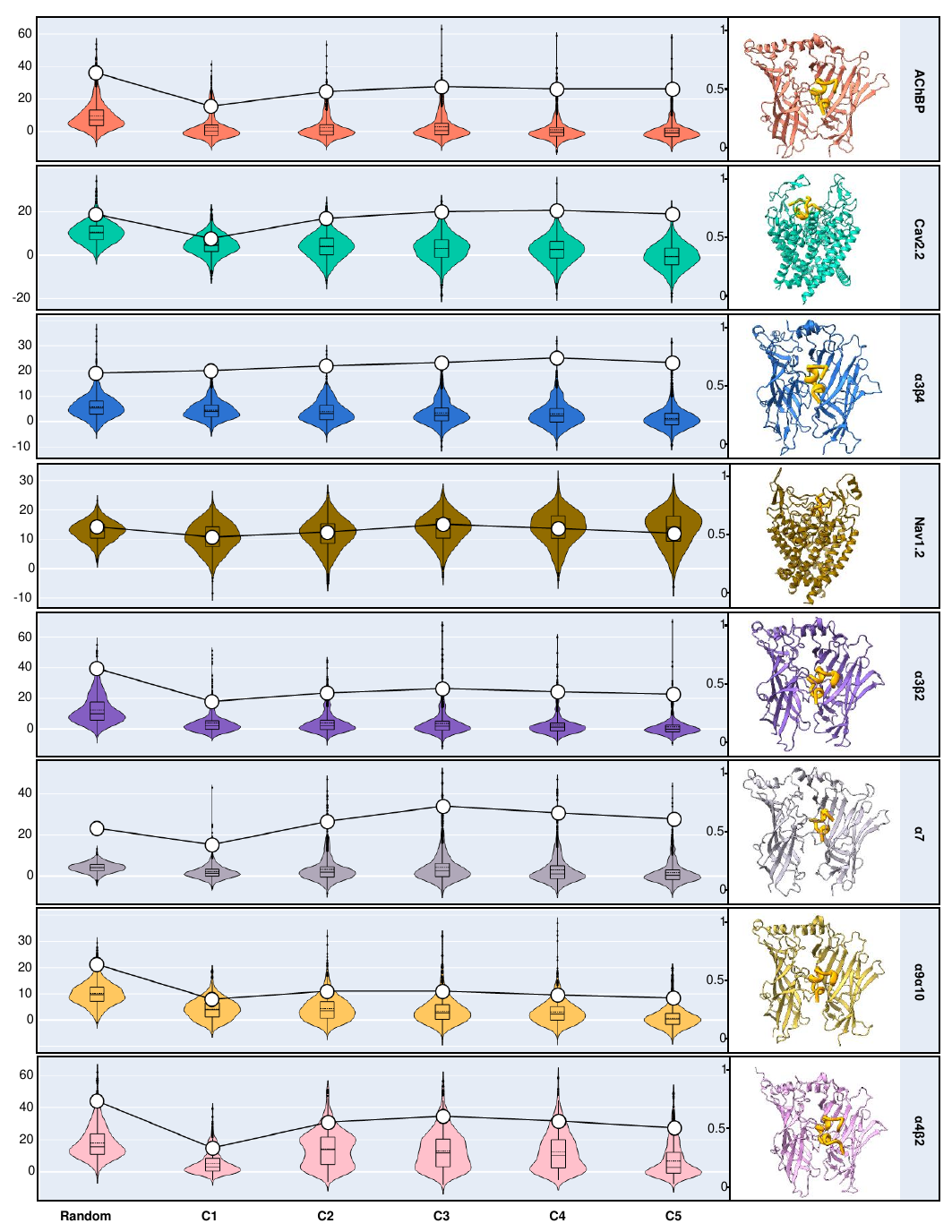}
\caption{\textbf{Performance evaluation of the data augmentation pipeline.} The x-axis represents different stages of random mutation and data augmentation, while the y-axis shows the $\Delta\Delta G$ values calculated using FoldX. Violin plots display the distribution of $\Delta\Delta G$ values for 1,000 peptide mutants, with the median value represented by the black horizontal line and a dashed horizontal line representing the mean within the box, the lower and upper quartiles delineating the borders of the box, and the vertical black lines indicating the 1.5 interquartile range. A line graph overlays the plots to indicate the average hamming distance between each mutant and the wild-type peptide. For each plot, the left y-axis represents $\Delta\Delta G$ values, the right y-axis corresponds to Hamming distance, and the right panel shows the representative complex structures for each system.
}\label{fig3}
\end{figure}

\subsection*{Improving CreoPep generation performance with data augmentation}
To train CreoPep, we curated 2,088 conotoxins with experimentally determined targets and potency values (IC\textsubscript{50}) from the ConoServer database \cite{ref8}, encompassing seven receptor families divided into 53 subtypes. Potency was classified as high (IC\textsubscript{50} $\leq$ \SI{1000}{\nano M}) or low (IC\textsubscript{50} > \SI{1000}{\nano M}). However, the dataset exhibited significant sparsity and redundancy; for example, only 67 high-potency conotoxins targeted the $\alpha$7 nAChR subtype, and most of these were mutants derived from a small number of wild-type sequences via residue scanning, leading to low sequence diversity. To address these limitations, we developed an iterative augmentation pipeline that integrates CreoPep with FoldX (Fig.~\ref{fig2}a, c), enabling progressive expansion of the training dataset through \textit{in silico} generation and energy-based filtering while preserving pharmacological relevance.

FoldX was selected for its well-established accuracy in predicting binding free energy changes ($\Delta\Delta G$) \cite{ref41}, and its generalizability was validated across seven conotoxin-receptor systems using experimental IC\textsubscript{50} data from ConoServer (TABLE S1--S14). As summarized in TABLE S15, FoldX achieved >75\% accuracy in five systems, with lower performance in $\alpha$-conotoxin Vc1.1/$\alpha$9$\alpha$10 and ImI/$\alpha$7 (<70\% accuracy). Notably, the ImI/$\alpha$7 system exhibited \SI{100}{\%} sensitivity, correctly identifying all potency-enhancing mutants. Comparable predictive performance for $\mu$-conotoxin KIIIA/Nav1.2 and $\alpha$-conotoxin LvIA/AChBP, while exceptional specificity (100\%) was achieved in the $\omega$-conotoxin MVIIA/Cav2.2 and $\alpha$-conotoxin AuIB/$\alpha$3$\beta$4 systems. Overall predictive reliability was further supported by Matthews Correlation Coefficient (MCC) analysis, with the highest performance recorded in the LvIA/AChBP system (MCC = 0.83). These results confirm FoldX as a selective and robust screening filter for high-potency variants in our pipeline.

\begin{figure}[t!]
\centering
\includegraphics[width=0.9\textwidth]{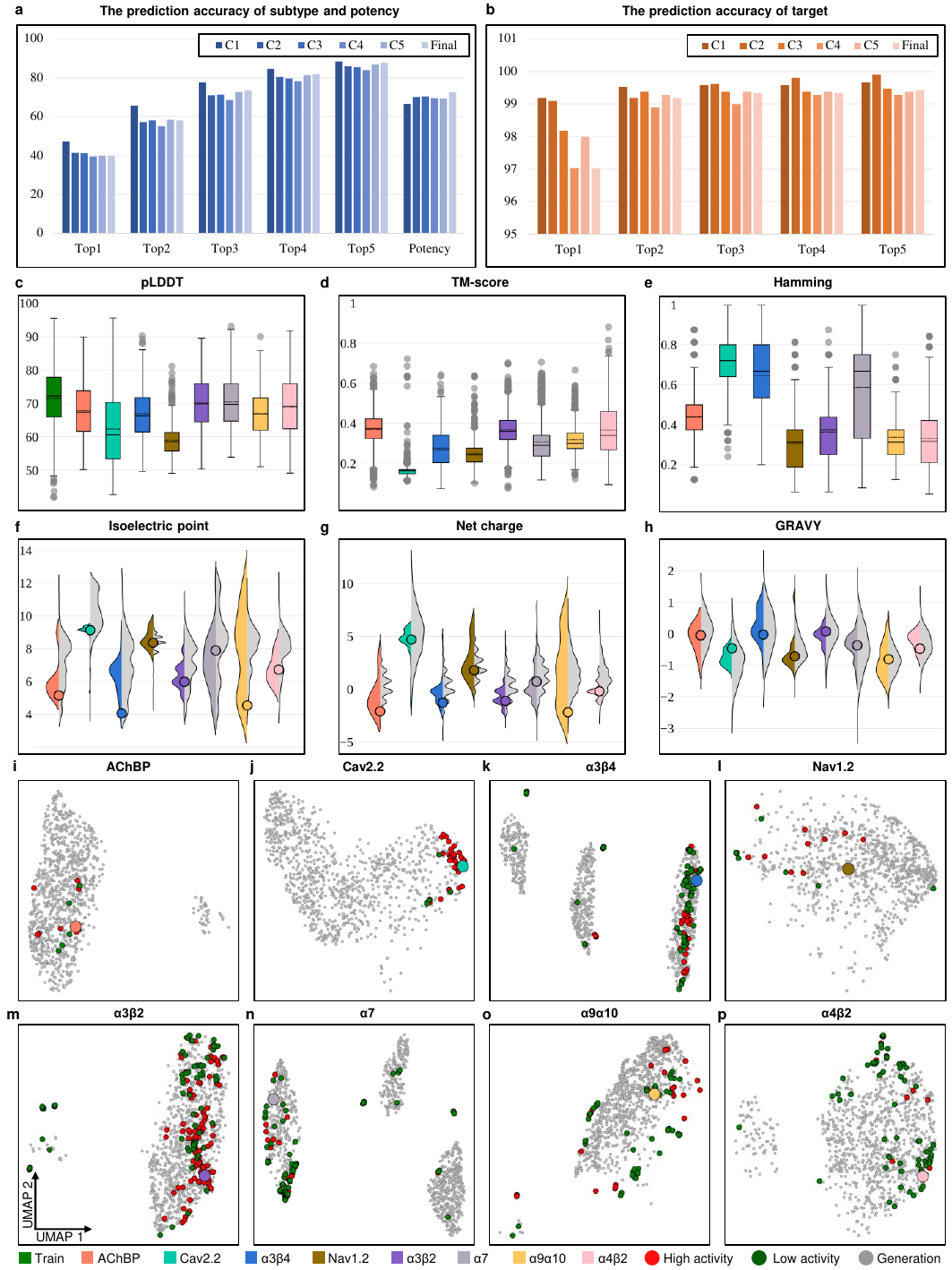}
\caption{\textbf{Classification and generative performance assessment of CreoPep.} \textbf{a}, Accuracy of CreoPep in subtype and potency prediction tasks. \textbf{b}, Accuracy of CreoPep in target prediction. \textbf{c-e}, Average pLDDT score, TM-score, and Hamming distance of 1,000 mutants generated by CreoPep. \textbf{f-h}, Comparison of isoelectric point, net charge, GRAVY (hydrophobicity) distributions among the wild-type conotoxin, high potency conotoxins, and 1,000 CreoPep-generated mutants. \textbf{i-p}, Feature distributions in the latent space for the wild-type conotoxin, high potency conotoxins, and 1,000 CreoPep-generated mutants across eight systems.
}\label{fig4}
\end{figure}

CreoPep was initially trained on the 2,088 conotoxin sequence dataset, then used to generate 1,000 mutant variants for each of eight distinct conotoxin/receptor subtype systems. To control the diversity of generated sequences, a polynomial sampling strategy with a temperature factor ($\tau$) was applied. In the first generation round, $\tau$ was set to 1.5 to maximize peptide sequence diversity. FoldX then computed $\Delta\Delta G$ values between generated mutants and their corresponding wild-type peptides. Mutants with $\Delta\Delta G \leq$ \SI{0.5}{\kilo\calorie\per\mole} were retained as high potency candidates, while sequences with $\Delta G$ $\geq$ \SI{0}{\kilo\calorie\per\mole} were labelled low potency (see Methods). Both sets were merged into the training pool to retrain CreoPep. This procedure was repeated over five rounds, with $\tau$ gradually reduced from 1.5 to 1.1 to refine the search toward higher-fidelity variants. The final augmented training set consisted of 18,355 conotoxins (Supplementary File 2), significantly enhancing model performance and sequence diversity while maintaining pharmacological relevance.

\subsection*{Design of high-potency conotoxin mutants with controllable diversity via CreoPep}
The high sequence conservation and limited number of potent conotoxin mutants in the training set constrain the diversity of generated sequences. However, excessive diversification may compromise potency. To address this trade-off, we implemented and evaluated a data augmentation pipeline designed to balance potency and diversity in the generated conotoxin mutants.

We calculated $\Delta\Delta G$ values for 1,000 mutants across eight systems at various stages of the augmentation process (Fig.~\ref{fig3}, violin plot). From stages C1 to C5, the $\Delta\Delta G$ distributions in most systems progressively shifted toward lower values, indicating an increased frequency of high-potency conotoxins ($\Delta\Delta G \leq$ \SI{0.5}{\kilo\calorie\per\mole}). In contrast, the KIIIA/Nav1.2 system exhibited a shift toward higher $\Delta\Delta G$ values, suggesting reduced efficacy in potency optimization. Nonetheless, the system still yielded approximately 20 favorable mutants ($\Delta\Delta G \leq$ \SI{0.5}{\kilo\calorie\per\mole}), outperforming random generation (TABLE S16).

To quantify sequence diversity, we calculated the Hamming distance, a measure of sequence divergence, (range: 0-1) between each mutant and their corresponding wild-type peptide (Fig.~\ref{fig3}, line plot). From stages C1 to C3, the average Hamming distance increased across all systems, reflecting a model bias toward exploring sequence diversity. Beyond C3, this trend reversed, except for minor fluctuations in MVIIA/Cav2.2 and AuIB/$\alpha$3$\beta$4 systems, indicating a shift toward potency optimization.

Importantly, the temperature factor ($\tau$) was initially set to 1.5 at stage C1 and gradually reduced to 1.1 by C5. Despite this reduction, the Hamming distance did not decline immediately (C1-C3), likely due to the incorporation of highly diverse mutants generated in C1 were incorporated into subsequent training cycles, thereby amplifying overall dataset diversity. During C1-C3, this cumulative diversity outweighed the decreasing influence of $\tau$. From C3–C5, however, the model exhibited a smooth transition from diversity exploration to potency refinement. Overall, these results demonstrate that CreoPep’s data augmentation pipeline effectively increases the number of high-potency mutants while maintaining enhanced sequence diversity.

\subsection*{Classification and generative performance assessment of CreoPep}
To evaluate the classification performance of CreoPep, we assessed the model’s prediction accuracy for subtype, potency, and target at each stage of data augmentation using the validation dataset. Given their unique pharmacological profiles, conotoxins can exhibit multitarget potency. Consequently, the model must be capable of recognizing and outputting multiple potential labels rather than being constrained to a single optimal prediction. To enable a comprehensive performance evaluation, we used the TopK accuracy metric, where a prediction is considered correct if the true label is among the top K predicted labels with the highest probabilities. As shown in Fig.~\ref{fig4}a, the model’s TopK accuracy for subtype prediction improved consistently with successive data augmentation iterations. By the final stage, the model achieved its highest Top1 to Top5 accuracy of \SI{76.63}{\%}, \SI{90.85}{\%}, \SI{95.92}{\%}, \SI{96.73}{\%}, and \SI{97.22}{\%}, respectively, significantly outperforming those at stage C1. A similar trend was observed for potency prediction, where the final stage reached a peak accuracy of \SI{93.36}{\%}, markedly higher than earlier stages (TABLE S17). These results confirm that iterative data augmentation enhances the model’s ability to accurately predict the conotoxin potency. Fig.~\ref{fig4}b further illustrates the model’s excellent performance in target prediction. From C1 to the final stage, Top1 to Top5 accuracy for target prediction remained consistently high. In the final stage, accuracy either matched or slightly exceeded earlier levels, achieving \SI{97.71}{\%} for Top1 and \SI{99.95}{\%} for Top5 (TABLE S18). This indicates that the model was already capable of high target prediction accuracy at early stages, and this performance was further refined with continued training.

To assess the model’s generative performance, we independently generated 1,000 conotoxin mutants for each of the eight subtype systems using the final model, and evaluated their foldability, novelty, physicochemical properties, and latent-space features. Structural foldability, predicted using OmegaFold \cite{ref42}, yielded average predicted Local Distance Difference Test (pLDDT) scores \cite{ref43} ranging from 58.93 (Nav1.2) to 70.43 ($\alpha$7 nAChR) (Fig.~\ref{fig4}c), suggesting reasonable structural viability, though slightly below the training set average of 71.46. Novelty analysis showed that the generated mutants exhibited low structural similarity to wild-type peptides (Template Modeling (TM) score \cite{ref44} < 0.5, Fig.~\ref{fig4}d) and high sequence divergence (Hamming distance > 0.3, Fig.~\ref{fig4}e). Mutants derived from $\mu$-conotoxin MVIIA/Cav2.2 demonstrated especially high novelty with TM-scores < 0.2 and Hamming distance > 0.7. Analysis of physicochemical properties revealed that the generated mutants retained essential features of their wild-type counterparts, including isoelectric point (Fig.~\ref{fig4}f), net charge (Fig.~\ref{fig4}g), and the Grand Average of Hydropathy (GRAVY) index \cite{ref45} (Fig.~\ref{fig4}h), with distributions closely matching those of high-potency training set peptides. Consistent trends were observed across other data augmentation cycles (Fig. S1). Finally, embeddings generated using Evolutionary Scale Modeling-2 (ESM-2 \cite{ref46}) and visualized through Uniform Manifold Approximation and Projection (UMAP \cite{ref47}) visualization (Fig.~\ref{fig4}i-p) confirmed strong latent-space overlap between the generated and wild-type peptides, particularly for the $\alpha$3$\beta$4, $\alpha$3$\beta$2, and $\alpha$9$\alpha$10 nAChR subtypes. This suggests that the model preserves evolutionary and structural features in its generative outputs. We also compared the latent-space distribution of mutants generated via random saturation mutagenesis to those produced during various augmentation stages (Fig. S2). Randomly saturated mutants showed minimal overlap with the training set peptides, while the augmented mutants initially clustered with the training data (C1) and then gradually dispersed into novel regions, forming distinct clusters. Notably, after C3, a subset of mutants reconverged with the training set cluster, consistent with the trends observed in Fig.~\ref{fig3}. Taken together, these findings demonstrate that our model is capable of generating structurally viable and novel conotoxin analogues that preserve key physicochemical and functional characteristics of the wild-type peptides.

\begin{figure}[t!]
\centering
\includegraphics[width=0.9\textwidth]{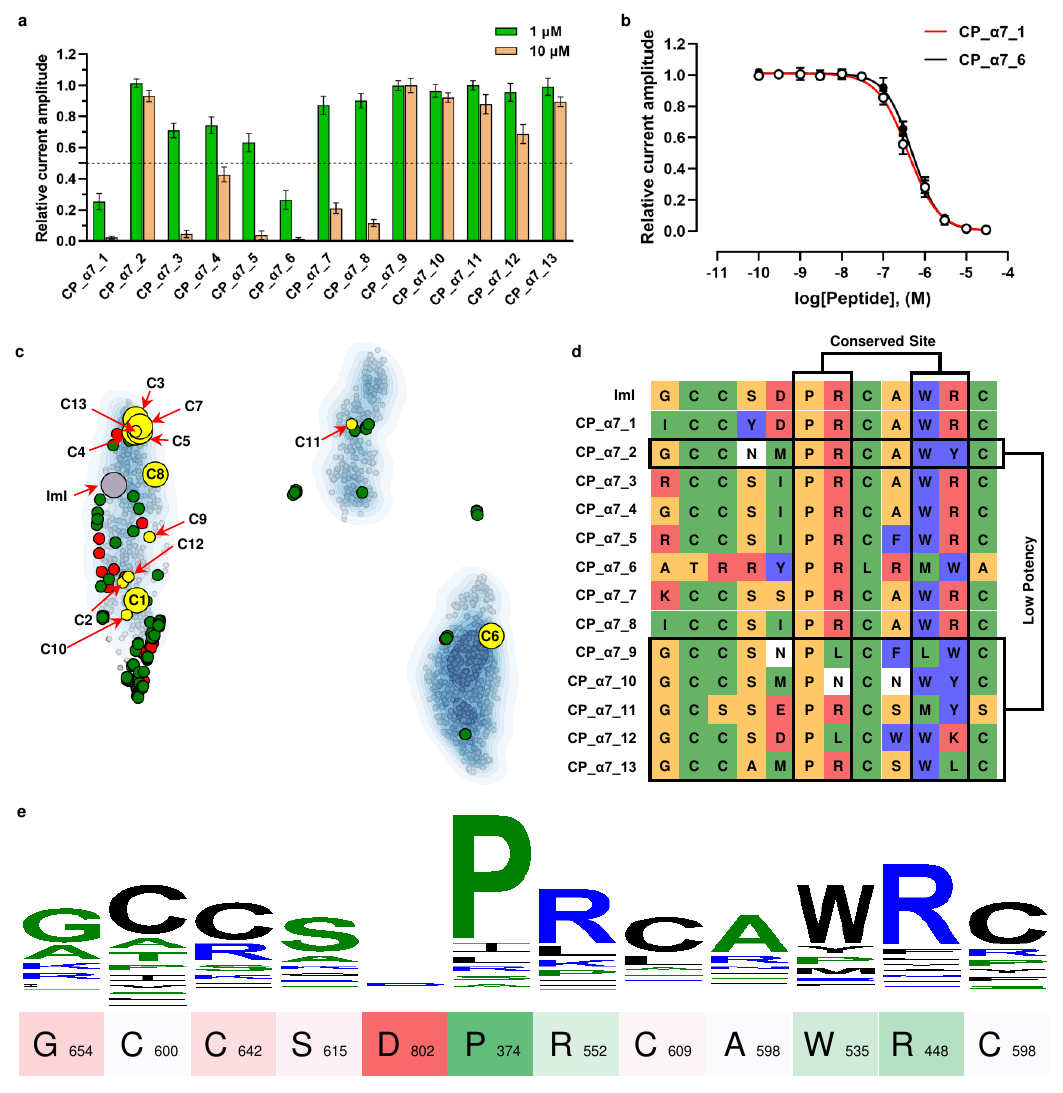}
\caption{\textbf{Classification and generative performance assessment of CreoPep.} \textbf{a}, Bar graph showing the inhibitory effects of 13 candidate conotoxins (at 1 and 10 µM) on ACh-evoked peak current amplitudes mediated by human (h) $\alpha$7 nAChRs. Whole-cell currents were evoked by 100 $\mu$M ACh (mean $\pm$ SD, $n$ = 4-8). The dashed line indicates 50\% inhibition of the peak current amplitude. \textbf{b}, Concentration-response relationships for CreoPep-$\alpha$7-1 and CreoPep-$\alpha$7-6 inhibition of ACh-evoked currents at (h) $\alpha$7 nAChRs. Current amplitudes (mean $\pm$ SD; $n$ = 6) were normalized to the response elicited by 100 $\mu$M ACh alone. \textbf{c}, Latent space distribution of 1,000 mutants (gray dots), ImI, high potency $\alpha$7-targeting conotoxins (red dots), low potency $\alpha$7-targeting conotoxins (green dots), and the 13 candidate conotoxins (large yellow dots for high potency; small yellow dots for low potency). The contour plot represents the density distribution of the 1,000 mutants, with darker colors indicating higher density in that region. \textbf{d}, Multiple sequence alignment of ImI with the 13 candidate conotoxins. \textbf{e}, Graphical representation of multiple sequence alignment of 1,000 CreoPep-generated mutants, showing the frequency of amino acid variations at each position relative to the ImI sequence.
}\label{fig5}
\end{figure}

\subsection*{Experimental validation of conotoxin ImI mutants generated via CreoPep}
We trained the final version of CreoPep on 18,355 conotoxins generated through our data augmentation pipeline. Focusing on the ImI/$\alpha$7 nAChR system selected for its experimental tractability, we generated 1,000 candidate mutants under high-potency conditions ($\tau$ = 1). Candidates were initially ranked by their predicted potency probabilities, and the top 60 candidates were selected for structural analysis. For each peptide-$\alpha$7 complex, five confirmations were predicted using AlphaFold3, followed by binding energy calculations with FoldX. Fifteen mutants consistently exhibited stronger binding than the wild-type ImI. Visual inspection of their binding modes in PyMOL further refined this list to 13 high-confidence candidates (CP\_$\alpha$7\_1 to CP\_$\alpha$7\_13), which were selected for experimental validation.

Using solid-phase peptide synthesis, we successfully produced all 13 candidate conotoxins for functional characterization. Two-electrode voltage clamp electrophysiology on human (h) $\alpha$7 nAChR,  revealed that seven peptides showed substantial inhibitory activity (>50\% blockade) at \SI{10}{\micro M} (Fig.~\ref{fig5}a). Notably, two designed peptides, CP\_$\alpha$7\_1 and CP\_$\alpha$7\_6, retained strong inhibition even at \SI{1}{\micro M}. Quantitative concentration-response analysis confirmed submicromolar IC\textsubscript{50} values for these top candidates, 405.1 $\pm$ \SI{28.7}{\nano M} for CP\_$\alpha$7\_1 and 504.6 $\pm$ \SI{29.8}{\nano M} for CP\_$\alpha$7\_6 (mean $\pm$ SD, n=6) (Fig.~\ref{fig5}b), validating CreoPep’s effectiveness in generating highly potent nAChR antagonists (Supplementary File 3).

To evaluate CreoPep’s capacity for exploring chemical space,  we mapped the 13 candidate conotoxin mutants within the latent space of the ImI/$\alpha$7 system. The mutants formed three distinct and well-separated clusters (Fig.~\ref{fig5}c). Interestingly, most candidate conotoxins co-localized with ImI, with four high-potency peptides forming a tightly packed subcluster. Notably, CP\_$\alpha$7\_6 was located in a region distant from the main cluster, in a densely populated area of the latent space. These findings highlight CreoPep’s ability  to explore chemical space and discover novel, high potency peptide variants. An integrated computational-experimental analysis demonstrated CreoPep’s robust capacity to identify functional residues in target-specific peptides. Multiple sequence alignment (Fig.~\ref{fig5}d) and variation frequency analysis (Fig.~\ref{fig5}e) identified four conserved non-cysteine residues (P6, R7, W10, and R11) as essential for activity. These residues were preserved across all high-potency mutants except CP\_$\alpha$7\_6; their mutation consistently abolished potency. Conversely, mutation-tolerant regions clustered at residues 1, 4, 5, and 9 (Fig.~\ref{fig5}e). All active mutants included substitutions at these locations, with position D5 displaying the highest mutational flexibility, six of seven potent variants carried modifications at this site. Furthermore, comparative analysis revealed significantly greater disulfide bond variability in the ImI/$\alpha$7 system compared to seven other systems (Fig. S3). This observation challenges the traditional view that disulfide bonds are rigid and conserved structural elements essential for $\alpha$-conotoxin bioactivity. However, it aligns with Tabassum \textit{et al.} \cite{ref48} who reported that, except for ImI, disulfide-deficient variants of Vc1.1 and AuIB showed diminished or abolished nAChR inhibitory potency. The strong concordance between CreoPep’s predictions and experimental results underscores its exceptional ability to identify critical functional residues, distinguish non-essential positions, accurately model mutation tolerance, and guide rational peptide design through targeted modification of variable regions while preserving core functional motifs.

\subsection*{Structural Basis of High-Potency Mutant Binding}
\begin{figure}[t!]
\centering
\includegraphics[width=0.9\textwidth]{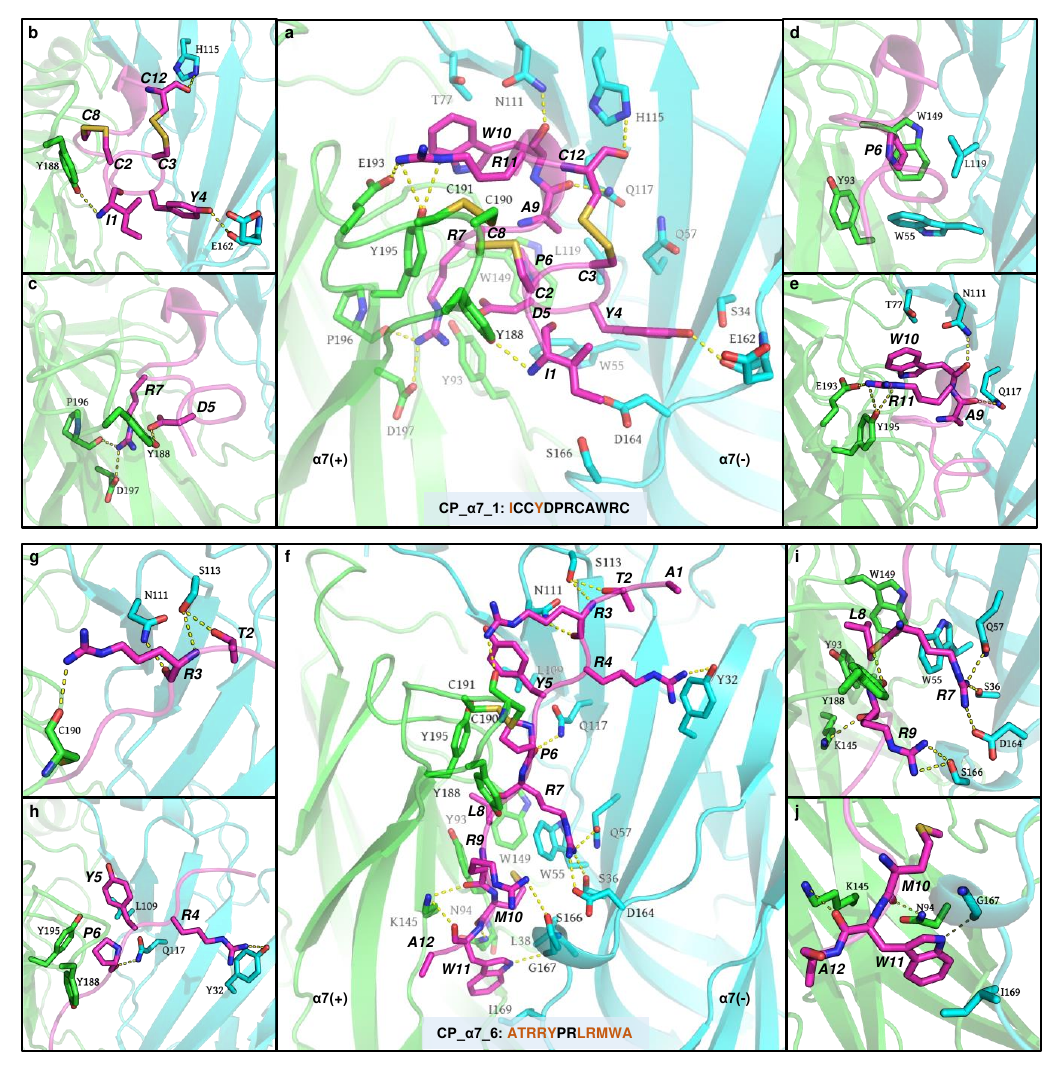}
\caption{\textbf{Constructed complex models of CP-$\alpha$7-1 and CP-$\alpha$7-6 bound to h$\alpha$7 nAChR and their binding modes.} \textbf{a}, Visualization of the overall binding mode between CP\_$\alpha$7\_1 and the h$\alpha$7 nAChR. \textbf{b-e}, Detailed interactions of individual amino acid residues of CP\_$\alpha$7\_1 with the h$\alpha$7 nAChR. \textbf{f}, Visualization of the binding mode between CP\_$\alpha$7\_6 and the h$\alpha$7 nAChR. \textbf{g-j}, Detailed interaction of individual amino acid residues of CP\_$\alpha$7\_6 with the h$\alpha$7 nAChR binding pocket.
}\label{fig6}
\end{figure}

Electrophysiological characterization identified CP\_$\alpha$7\_1 and CP\_$\alpha$7\_6 as the most potent h$\alpha$7 nAChR inhibitors among the CreoPep-designed variants. AlphaFold3-predicted structures, further refined by MD simulations, revealed that both mutants employ distinct strategies to stabilize the $\alpha$7(+)$\alpha$7(-) interface. CP\_$\alpha$7\_1 preserves ImI's disulfide bonds (Cys2-Cys8, Cys3-Cys14) and core binding motifs, whereas CP\_$\alpha$7\_6 achieves comparable potency via a completely reengineered interaction network, despite lacking disulfide connectivity (Fig.~\ref{fig5}d). In the CP\_$\alpha$7\_1/h$\alpha$7 nAChR complex, key stabilizing interactions include hydrogen bonds (I1-Y188, Y4-E162, D5-Y188, A9-Q117), hydrophobic packing (P6 with Y93/W149/W55/L119), and salt bridges (R7-D197, R11-E193) (Fig.~\ref{fig6}a-e). In contrast, CP\_$\alpha$7\_6-ha7 complex features nine hydrogen bonds (spanning T2-S113 to A12-K145), hydrophobic contacts involving Y5-L109, P6-Y188/Y195, L8-W149, and W11-L169, as well as a critical salt-bridge (R7-D164) supported by additional hydrogen bonds (R7-S36/Q57) (Fig.~\ref{fig6}f-j). Despite substantial sequence divergence, both variants achieve interfacial stabilization through distinct interaction networks. This demonstrates CreoPep’s capability to uncover both conservative and non-conservative binding modes without sacrificing potency, highlighting its potential for rational peptide engineering.

\section*{Discussion}
Target-specific peptides hold great promise in drug development due to their high affinity and specificity. However, the clinical translation of natural peptides is often limited by insufficient activity, poor selectivity, and low stability. Traditional approaches, such as saturation and point mutagenesis, have partially addressed the challenge of low potency, but these methods remain inefficient and heavily dependent on expert knowledge. Therefore, there is a critical need for a universal and efficient strategy to design potent, target-specific peptides with high sequence diversity. 

In this study, we present CreoPep, a deep learning-based conditional generation framework that leverages a masked language model with a Progressive Masking (PM) strategy for efficient peptide design and optimization. The key innovation of CreoPep lies in its ability to generate diverse peptides guided by target and potency labels, while maintaining sequence diversity and biological functionality through a temperature-controlled multinomial sampling approach. Furthermore, CreoPep integrates FoldX-based binding energy screening into an iterative data augmentation pipeline, enriching the training dataset, improving model performance, and offering valuable design insights particularly under few-shot learning conditions. Notably, CreoPep serves as an efficient computational alternative to traditional amino acid scanning for identifying critical functional residues. Electrophysiological validation demonstrated the model’s strong performance: 7 out of 13 designed conotoxin candidates showed relatively high potency against the h$\alpha$7 nAChR. Among them, the variants CP\_$\alpha$7\_1 and CP\_$\alpha$7\_6 exhibited sub-micromolar inhibitory activity, with IC$_{50}$ values of \SI{405.1}{\nano M} and \SI{504.6}{\nano M} , respectively. These findings highlight CreoPep’s ability to generate potent peptide mutants with therapeutic potential.

Despite its strengths, CreoPep has several limitations. For instance, among the ImI conotoxin mutants generated, only one variant exhibited marginal improvement in potency compared to the wild-type peptide. This outcome highlights the inherent difficulty of enhancing ImI, a naturally optimized short peptide. To address this limitation, the inclusion of non-natural modifications could offer an effective path forward. Currently, CreoPep is limited the use of natural amino acids, thereby a broader chemical space that could further enhance therapeutic properties. In conclusion, CreoPep represents a robust, deep learning-driven platform for the rational design of target-specific peptide therapeutics. Future developments could focus on three main directions: (1) incorporating non-natural amino acids to access unexplored chemical space, (2) designing multi-target peptides for complex disease networks, and (3) integrating patient-specific biological data to enable personalized therapeutic strategies. These advancements would further establish CreoPep’s role in advancing next-generation peptide drug discovery.

\section*{Methods}
\subsection*{Data collection and processing}
We collected 2,088 conotoxins from the ConoServer database \cite{ref8}, with each entry containing target information and potency data. Each training sample required a subtype label, a potency label, a wild-type conotoxin, and up to three auxiliary conotoxins. The auxiliary conotoxins are randomly selected from a set which has the same subtype and potency as the wild-type conotoxin. Additionally, we performed data augmentation across eight complex systems, assigning pseudo-labels based on binding energy calculated by FoldX. The expanded datasets from each round contained 4,471, 7,330, 10,499, 14,072, and 18,355 peptide sequences, respectively.

\subsection*{Model implementations}
The CreoPep model architecture integrates four core components: the Encoder Block (for labels and wild-type conotoxin sequence representation learning), the MLP Block (for processing auxiliary conotoxin features), the CNN Block (for dimensionality reduction after feature fusion), and the Decoder Block (for final prediction). CreoPep first extracts wild-type conotoxin sequence, as well as subtype and potency labels, using features through the Encoder Block. This Block is based on the pre-trained protein language model ProtBert. The BERT encoder of ProtBert is frozen, and an additional trainable BERT encoder without an embedding layer is added, enabling the model to adapt to specific tasks while leveraging pre-trained knowledge. The MLP Block integrates an embedding layer and a Xavier-initialized multilayer perceptron (Linear layer + LeakyReLU + Linear layer) to extract auxiliary conotoxin features. Subsequently, these features are concatenated and then reduced in dimension by a $1\times1$ convolutional CNN block.Finally, the fused features are input into the Decoder Block to generate the final prediction. The pseudocode of the core training process is shown in Supplementary Algorithm 1.

\subsection*{Loss Function}
We propose a masked cross-entropy loss function to handle multi-task token prediction. For a sequence with $n$ positions, the raw cross-entropy loss $\mathcal{L}_{\mathrm{CE}}(y_i, \hat{y}_i)$ is computed at each position $i$. We then apply three binary mask matrices -- the \texttt{[MASK]} token position mask $\mathbb{M}_{\mathrm{mask}} \in \{0,1\}^n$, label mask $\mathbb{M}_{\mathrm{label}} \in \{0,1\}^m$, and sequence mask $\mathbb{M}_{\mathrm{seq}} \in \{0,1\}^p$ -- to compute task-specific losses:

\begin{align}
\mathcal{L}_{\mathrm{mask}} &= \frac{
    \sum\limits_{i=1}^n \bigl(\mathcal{L}_{\mathrm{CE}}(y_i, \hat{y}_i) \circ \mathbb{M}_{\mathrm{mask},i}\bigr)
}{
    \sum\limits_{i=1}^n \mathbb{M}_{\mathrm{mask},i}
} \label{eq:position_loss} \\
\mathcal{L}_{\mathrm{label}} &= \frac{
    \sum\limits_{j=1}^m \bigl(\mathcal{L}_{\mathrm{CE}}(y_j, \hat{y}_j) \circ \mathbb{M}_{\mathrm{label},j}\bigr)
}{
    \sum\limits_{j=1}^m \mathbb{M}_{\mathrm{label},j}
} \label{eq:label_loss} \\
\mathcal{L}_{\mathrm{seq}} &= \frac{
    \sum\limits_{k=1}^p \bigl(\mathcal{L}_{\mathrm{CE}}(y_k, \hat{y}_k) \circ \mathbb{M}_{\mathrm{seq},k}\bigr)
}{
    \sum\limits_{k=1}^p \mathbb{M}_{\mathrm{seq},k}
} \label{eq:sequence_loss}
\end{align}

where $\circ$ denotes element-wise multiplication (Hadamard product). The final composite loss is given by the weighted sum:

\begin{equation}
\mathcal{L}_{\text{total}} = \alpha\mathcal{L}_{\mathrm{pos}} + \beta\mathcal{L}_{\mathrm{label}} + \gamma\mathcal{L}_{\mathrm{seq}}
\label{eq:total_loss}
\end{equation}

Here, $\alpha,\beta,\gamma \in \mathbb{R}^+$ are tunable hyperparameters that control each task’s contribution to the overall optimization objective, with $\alpha$=1, $\beta$=1/2, and $\gamma$=1/3. This loss function ensures that the model prioritizes position-level correctness while also incorporating label and sequence-level constraints with reduced weighting.

\subsection*{Mask ratio setting}
To evaluate the effect of different masking ratios on the generative model, we tested values ranging from 10\% to 100\% in 10\% increments. All sequences were first padded to a fixed length of $L=54$ tokens. For each masking ratio $r \in \{0.1, 0.2, \ldots, 1.0\}$, we computed the number of masked positions as:

\begin{equation}
t = r \times L
\end{equation}

where t is computed using rounding to the nearest integer and $t=54$ represents complete masking (100\%) and $t=27$ corresponds to the 50\% masking condition. Under a fixed 100-epoch training regime, all model variants achieved convergence as shown in Fig. S4, with asterisks marking their respective minimum loss values. The optimal performance was observed at $r=0.5$ (50\% masking, $t=27$ positions), yielding the lowest loss of 0.729. This 50\% masking configuration was consequently adopted for final model training.

\subsection*{Sampling strategy}
In the data augmentation pipeline, diversity in the generated sequences is ensured by introducing multinomial sampling during the peptide generation phase. Meanwhile, a temperature factor $\tau$ is incorporated into the sampling process to control the degree of diversity in the generated peptides. Specifically, a higher $\tau$ value results in a smoother probability distribution, thereby increasing the diversity of the sampled peptides. The calculation formula is as follows:

\begin{equation}
{P(y_i)} = \frac{\exp(z_i/\tau)}{\sum\limits_{j=1}^{V} \exp(z_j/\tau)}
\label{eq:temperature_scaling}
\end{equation}

Here, $z_i$ denotes the logit (raw score) for the $i$-th class, while $z_j$ represents the logit for all $V$ classes, with $j$ ranging from $1$ to $V$. The output $P(y_i)$ is the resulting probability distribution. Subsequently, a class is sampled from this distribution using multinomial sampling:

\begin{equation}
\hat{x} \sim \mathrm{Multinomial}\bigl(P(y_1), P(y_2), \ldots, P(y_V)\bigr)
\label{eq:multinomial_sampling}
\end{equation}

\subsection*{Evaluation metrics for FoldX}
We evaluated FoldX's performance using four standard metrics: accuracy, sensitivity, specificity, and MCC (Matthews correlation coefficient), with their calculation formulas shown as follows. 

\begin{equation}
\text{Accuracy} = \frac{TN + TP}{TN + TP + FN + FP}
\label{eq:accuracy}
\end{equation}

\begin{equation}
\text{Sensitivity} = \frac{TP}{TP + FN}
\label{eq:sensitivity}
\end{equation}

\begin{equation}
\text{Specificity} = \frac{TN}{TN + FP}
\label{eq:specificity}
\end{equation}

\begin{equation}
\text{MCC} = \frac{TP \times TN - FP \times FN}{\sqrt{(TP + FP)(TP + FN)(TN + FP)(TN + FN)}}
\label{eq:mcc}
\end{equation}

where TP, TN, FP and FN denote the number of true positives, true negatives, false positives and false negatives, respectively. The complex structure that achieved optimal performance across all four evaluation metrics was selected as the final system structure. Although $\alpha$-conotoxin GID contains non-natural amino acids that are incompatible with FoldX's input format, we included it in subsequent data augmentation tasks due to the importance of the GID/$\alpha$4$\beta$2 system. For data augmentation, we used GID[(GLA)4E, O16P] as the representative complex structure of the GID/$\alpha$4$\beta$2 system, but we did not assess FoldX's performance on this particular complex.

\subsection*{Binding Energy Calculations}
We utilized official Python bindings of FoldX (\texttt{pyfoldx}) to calculate peptide-receptor binding free energies. Using the wild-type complex structure as a template, we first optimize the structure using FoldX's \texttt{repair} function for energy minimization. We then calculate the binding free energy of wild-type peptide ($\Delta G_{\text{wild-type}}$) using \texttt{getInterfaceEnergy} function. For each mutant, the following steps were executed: (i) mutation site identification through sequence alignment; (ii) stepwise introduction of amino acid substitutions; (iii) local energy optimization following each mutation; and (iv) global relaxation of the fully mutated complex. The binding free energy of each mutant ($\Delta G_{\text{mutant}}$) was then computed with \texttt{getInterfaceEnergy}, and the binding energy difference ($\Delta\Delta G = \Delta G_{\text{mutant}} - \Delta G_{\text{wild-type}}$) was determined. During dataset expansion, we established robust classification criteria based on prior studies. Following Wu et al.'s \cite{ref41} definition, in which $\Delta\Delta G = \pm\SI{1}{\kilo\calorie\per\mole}$ indicates comparable binding affinity, we adopted a more stringent threshold ($\Delta\Delta G \leq \SI{0.5}{\kilo\calorie\per\mole}$) to identify favorable mutations. These were labeled as high-potency mutants (positive samples) in our training dataset. For low-potency mutants (negative samples), we applied a conservative cutoff ($\Delta G_{\text{mutant}} > \SI{0}{\kilo\calorie\per\mole}$) to reduce the likelihood of false negatives and enhance dataset reliability.

\subsection*{Molecular dynamics simulations of mutants CP\_$\alpha$7\_1 and CP\_$\alpha$7\_6 with  h$\alpha$7 nAChR}
The initial complex structures of the CP\_$\alpha$7\_1 and CP\_$\alpha$7\_6 mutants bound to the h$\alpha$7 nAChR were predicted using AlphaFold3. Based on this structure, the C-terminal amidation modifications of the mutants were performed using PyMOL. MD simulations were performed with the AMBER~22 software package using ff19SB force field \cite{ref49}. Each complex was solvated in a truncated octahedral periodic box. The system was initially heated gradually from 50~K to 300~K over 100~ps in the NVT ensemble, with the solute restrained by a harmonic force of 5~kcal~mol$^{-1}$~\AA$^{-2}$. Subsequently, the simulation transitioned to the NPT ensemble, during which the harmonic restraints were gradually reduced from 5~kcal~mol$^{-1}$~\AA$^{-2}$ to 0~kcal~mol$^{-1}$~\AA$^{-2}$ over another 100~ps. For the production phase, a 50~ns MD simulation was carried out with the temperature and pressure maintained at 300~K and 1~bar, respectively. After MD simulation, MD trajectories were analyzed using VMD (\url{http://www.ks.uiuc.edu/}), and root mean square deviation (RMSD) values were calculated to assess structural stability.

\subsection*{Peptide Synthesis and Purification}
The linear peptide was synthesized on a 0.1 mmol scale using solid-phase peptide synthesis (SPPS) with Rink-Amide resin. After swelling and deprotecting the resin using 20\% piperidine for 2 h, Fmoc-protected amino acids were coupled using HCTU (4.0 equiv) and DIPEA (8.0 equiv) for 1 h, followed by deprotection with 20\% piperidine  for 30 min. Between each step, the resin was washed three times with DMF and twice with DCM. For peptides containing two disulfide bonds, regioselective oxidation was achieved by incorporating Fmoc-Cys(Acm)-OH at the Cys\textsuperscript{I} and Cys\textsuperscript{III} positions and Fmoc-Cys(Trt)-OH at Cys\textsuperscript{II} and Cys\textsuperscript{IV} positions. Following peptide chain assembly and terminal Fmoc removal, the peptide was cleaved from the resin using TFA/TIPS/H\textsubscript{2}O (90:5:5) for 3 h, concentrated, precipitated in cold ether, and analyzed by LC-MS. The first disulfide bond was formed by slowly adding 0.8 equiv of 2,2$'$-dithiodipyridine in 10 mL of methanol to the crude peptide solution over 30 min with constant constant stirring. The product was purified by preparative RP-HPLC. Subsequently, iodine (5 equiv) was used to simultaneously remove the Acm protecting groups and facilitate disulfide bond formation between Cys\textsuperscript{I} and Cys\textsuperscript{III}. The reaction was quenched with ascorbic acid until the solution became colorless. All crude peptides were purified by semi-preparative RP-HPLC on a C\textsubscript{18} column using a gradient of 0.05\% TFA in H\textsubscript{2}O/MeCN (flow rate: 6 mL/min) with the following gradient: 0 min, 100\% solvent A; 5 min, 85\% solvent A; 20 min, 70\% solvent A; 40 min, 60\% solvent A. Solvent A: H\textsubscript{2}O/MeCN (90:10, v/v); solvent B: H\textsubscript{2}O/MeCN (10:90, v/v). The purity and molecular weight of the final products were confirmed by analytical RP-HPLC and LC/MS, respectively. All peptides had a purity of greater than 95\% (Fig. S5-S17).

\subsection*{Xenopus laevis Oocyte Preparation and Microinjection}
All procedures were approved by the Animal Ethics Committees of the Victor Chang Cardiac Research Institute, Sydney, and the University of Wollongong (project no. AE 20/17). Human $\alpha7$ nAChR clones, provided by Prof. Jon Lindstrom (University of Pennsylvania) were linearized, and cRNA was transcribed \textit{in vitro} using the SP6/T7 mMessage mMachine kit (Ambion). Stage V--VI oocytes (1200--1300 $\mu$m) were harvested from 5-year-old female \textit{Xenopus laevis} (Nasco), anesthetized with 1.7 mg/mL ethyl 3-aminobenzoate methanesulfonate (pH 7.4), defolliculated with 1.5 mg/mL collagenase Type II (Worthington) in OR-2 solution (82.5 NaCl, 2 KCl, 1 MgCl$_2$, 5 HEPES, pH 7.4), and microinjected with 10 ng of h$\alpha7$ cRNA (verified by spectrophotometry and gel electrophoresis) using Drummond glass pipettes. Injected oocytes were incubated at 18$^\circ$C in ND96 solution (96 NaCl, 2 KCl, 1 CaCl$_2$, 1 MgCl$_2$, 5 HEPES, pH 7.4) supplemented with 5\% FBS, 50 mg/L gentamicin, and 10,000 U/mL penicillin--streptomycin (Gibco).

\subsection*{Oocyte Two-Electrode Voltage Clamp Recording and Data Analysis}
Electrophysiological recordings were performed 2--5 days post-injection using a GeneClamp 500B amplifier and pClamp9 software (Molecular Devices), with a holding potential of $-$80 mV. Recording electrodes (0.3--1 M$\Omega$ resistance) were pulled from GC150T-7.5 borosilicate glass (Harvard Apparatus) and filled with 3 M KCl. To reduce interference from Ca$^{2+}$-activated chloride channels, oocytes expressing h$\alpha7$ nAChR, were preincubated with 100 $\mu$M BAPTA-AM for $\sim$3 h prior to recording. Perfusion was carried out at 2 mL/min (Legato 270 syringe pump; KD Scientific) in an OPC-1 chamber ($<$20 $\mu$L; Automate Scientific) using ND96 solution. After rinsing, oocytes were exposed to three applications of 100 $\mu$M ACh (the EC$_{50}$ for $\alpha7$ nAChR), each followed by a 3 min washout with ND96. Peptide testing involved 5 min static incubation (perfusion off), followed by co-application with ACh during perfusion. All peptides were dissolved in ND96 supplemented with 0.1\% BSA. Current amplitudes (ACh alone vs. ACh + peptide) were analyzed using Clampfit 10.7 (Molecular Devices), with peptide activity expressed as the ratio of evoked currents. Data from 4--8 oocytes are presented as mean $\pm$ SD. Concentration--response curves were generated by nonlinear regression to determine IC$_{50}$ values (95\% CI), and statistical significance was assessed using ANOVA ($p < 0.05$; GraphPad Prism 9).

\section*{Data availability}
The model weights of CreoPep are available on Zenodo at this link (\url{https://zenodo.org/records/15192592}). The training sets, along with the generated peptides and their corresponding $\Delta\Delta G$ values from various data augmentation stages, can be accessed at this GitHub repository (\url{https://github.com/gc-js/CreoPep/tree/main/data}). Additionally, the source files and codes used for plotting are available at \url{https://github.com/gc-js/CreoPep/tree/main/plot}.

\section*{Code availability}
The code is available in our GitHub project at \url{https://github.com/gc-js/CreoPep}. We have also deployed the algorithm, including the model weights for CreoPep, on Hugging Face Spaces at \url{https://huggingface.co/spaces/oucgc1996/CreoPep}. 

\section*{Acknowledgments}
We gratefully acknowledge our laboratory members for their insightful discussions. This work was supported by the National Natural Science Foundation of China (Grant No. 82122064, 82473832), Jinan Innovation Team Project (202333022), Qingdao National Laboratory for Marine Science and Technology (2022QNLM030003-4), Shandong Supporting Funds for Talents (2022GJJLJRC02-046), National Natural Science Foundation of China (62373172), and an Australian Research Council Discovery Project Grant (DP150103990 to DJA).


\bibliographystyle{IEEEtran}  
\bibliography{references}

\begin{thebibliography}{10}
\providecommand{\url}[1]{#1}
\csname url@samestyle\endcsname
\providecommand{\newblock}{\relax}
\providecommand{\bibinfo}[2]{#2}
\providecommand{\BIBentrySTDinterwordspacing}{\spaceskip=0pt\relax}
\providecommand{\BIBentryALTinterwordstretchfactor}{4}
\providecommand{\BIBentryALTinterwordspacing}{\spaceskip=\fontdimen2\font plus
\BIBentryALTinterwordstretchfactor\fontdimen3\font minus \fontdimen4\font\relax}
\providecommand{\BIBforeignlanguage}[2]{{%
\expandafter\ifx\csname l@#1\endcsname\relax
\typeout{** WARNING: IEEEtran.bst: No hyphenation pattern has been}%
\typeout{** loaded for the language `#1'. Using the pattern for}%
\typeout{** the default language instead.}%
\else
\language=\csname l@#1\endcsname
\fi
#2}}
\providecommand{\BIBdecl}{\relax}
\BIBdecl

\bibitem{ref1}
M.~Muttenthaler, G.~F. King, D.~J. Adams, and P.~F. Alewood, ``Trends in peptide drug discovery,'' \emph{Nature Reviews Drug Discovery}, vol.~20, no.~4, pp. 309--325, 2021.

\bibitem{ref2}
L.~Wang, N.~Wang, W.~Zhang, X.~Cheng, Z.~Yan, G.~Shao, X.~Wang, R.~Wang, and C.~Fu, ``Therapeutic peptides: current applications and future directions,'' \emph{Signal Transduction and Targeted Therapy}, vol.~7, no.~1, p.~48, 2022.

\bibitem{ref3}
A.-H. Jin, M.~Muttenthaler, S.~Dutertre, S.~Himaya, Q.~Kaas, D.~J. Craik, R.~J. Lewis, and P.~F. Alewood, ``Conotoxins: chemistry and biology,'' \emph{Chemical Reviews}, vol. 119, no.~21, pp. 11\,510--11\,549, 2019.

\bibitem{ref4}
R.~J. Lewis, S.~Dutertre, I.~Vetter, and M.~J. Christie, ``Conus venom peptide pharmacology,'' \emph{Pharmacological Reviews}, vol.~64, no.~2, pp. 259--298, 2012.

\bibitem{ref5}
H.~Terlau and B.~M. Olivera, ``Conus venoms: a rich source of novel ion channel-targeted peptides,'' \emph{Physiological Reviews}, vol.~84, no.~1, pp. 41--68, 2004.

\bibitem{ref6}
S.~Pei, N.~Wang, Z.~Mei, D.~Zhangsun, D.~J. Craik, J.~M. McIntosh, X.~Zhu, and S.~Luo, ``Conotoxins targeting voltage-gated sodium ion channels,'' \emph{Pharmacological Reviews}, vol.~76, no.~5, pp. 828--845, 2024.

\bibitem{ref7}
K.~B. Akondi, M.~Muttenthaler, S.~Dutertre, Q.~Kaas, D.~J. Craik, R.~J. Lewis, and P.~F. Alewood, ``Discovery, synthesis, and structure--activity relationships of conotoxins,'' \emph{Chemical Reviews}, vol. 114, no.~11, pp. 5815--5847, 2014.

\bibitem{ref8}
Q.~Kaas, R.~Yu, A.-H. Jin, S.~Dutertre, and D.~J. Craik, ``Conoserver: updated content, knowledge, and discovery tools in the conopeptide database,'' \emph{Nucleic Acids Research}, vol.~40, no.~D1, pp. D325--D330, 2012.

\bibitem{ref9}
S.~Gao, X.~Yao, and N.~Yan, ``Structure of human cav2. 2 channel blocked by the painkiller ziconotide,'' \emph{Nature}, vol. 596, no. 7870, pp. 143--147, 2021.

\bibitem{ref10}
C.~M. Noviello, A.~Gharpure, N.~Mukhtasimova, R.~Cabuco, L.~Baxter, D.~Borek, S.~M. Sine, and R.~E. Hibbs, ``Structure and gating mechanism of the $\alpha$7 nicotinic acetylcholine receptor,'' \emph{Cell}, vol. 184, no.~8, pp. 2121--2134, 2021.

\bibitem{ref11}
X.~Pan, Z.~Li, X.~Huang, G.~Huang, S.~Gao, H.~Shen, L.~Liu, J.~Lei, and N.~Yan, ``Molecular basis for pore blockade of human na+ channel nav1. 2 by the $\mu$-conotoxin kiiia,'' \emph{Science}, vol. 363, no. 6433, pp. 1309--1313, 2019.

\bibitem{ref12}
L.~Wang, N.~Wang, W.~Zhang, X.~Cheng, Z.~Yan, G.~Shao, X.~Wang, R.~Wang, and C.~Fu, ``Therapeutic peptides: current applications and future directions,'' \emph{Signal Transduction and Targeted Therapy}, vol.~7, no.~1, p.~48, 2022.

\bibitem{ref13}
K.~Fosgerau and T.~Hoffmann, ``Peptide therapeutics: current status and future directions,'' \emph{Drug Discovery Today}, vol.~20, no.~1, pp. 122--128, 2015.

\bibitem{ref14}
B.~M. Olivera, L.~J. Cruz, V.~De~Santos, G.~LeCheminant, D.~Griffin, R.~Zeikus, J.~M. McIntosh, R.~Galyean, and J.~Varga, ``Neuronal calcium channel antagonists. discrimination between calcium channel subtypes using. omega.-conotoxin from conus magus venom,'' \emph{Biochemistry}, vol.~26, no.~8, pp. 2086--2090, 1987.

\bibitem{ref15}
G.~P. Miljanich, ``Ziconotide: neuronal calcium channel blocker for treating severe chronic pain,'' \emph{Current Medicinal Chemistry}, vol.~11, no.~23, pp. 3029--3040, 2004.

\bibitem{ref16}
P.~Kurtzhals, S.~{\O}stergaard, E.~Nishimura, and T.~Kjeldsen, ``Derivatization with fatty acids in peptide and protein drug discovery,'' \emph{Nature Reviews Drug Discovery}, vol.~22, no.~1, pp. 59--80, 2023.

\bibitem{ref17}
W.~Xiao, W.~Jiang, Z.~Chen, Y.~Huang, J.~Mao, W.~Zheng, Y.~Hu, and J.~Shi, ``Advance in peptide-based drug development: delivery platforms, therapeutics and vaccines,'' \emph{Signal Transduction and Targeted Therapy}, vol.~10, no.~1, p.~74, 2025.

\bibitem{ref18}
X.~Ji, A.~L. Nielsen, and C.~Heinis, ``Cyclic peptides for drug development,'' \emph{Angewandte Chemie}, vol. 136, no.~3, p. e202308251, 2024.

\bibitem{ref19}
M.~Baek, F.~DiMaio, I.~Anishchenko, J.~Dauparas, S.~Ovchinnikov, G.~R. Lee, J.~Wang, Q.~Cong, L.~N. Kinch, R.~D. Schaeffer \emph{et~al.}, ``Accurate prediction of protein structures and interactions using a three-track neural network,'' \emph{Science}, vol. 373, no. 6557, pp. 871--876, 2021.

\bibitem{ref20}
J.~Abramson, J.~Adler, J.~Dunger, R.~Evans, T.~Green, A.~Pritzel, O.~Ronneberger, L.~Willmore, A.~J. Ballard, J.~Bambrick \emph{et~al.}, ``Accurate structure prediction of biomolecular interactions with alphafold 3,'' \emph{Nature}, vol. 630, no. 8016, pp. 493--500, 2024.

\bibitem{ref21}
J.~Delgado, L.~G. Radusky, D.~Cianferoni, and L.~Serrano, ``Foldx 5.0: working with rna, small molecules and a new graphical interface,'' \emph{Bioinformatics}, vol.~35, no.~20, pp. 4168--4169, 2019.

\bibitem{ref22}
N.~London, B.~Raveh, E.~Cohen, G.~Fathi, and O.~Schueler-Furman, ``Rosetta flexpepdock web server—high resolution modeling of peptide--protein interactions,'' \emph{Nucleic Acids Research}, vol.~39, no. suppl\_2, pp. W249--W253, 2011.

\bibitem{ref23}
P.~Zhou, B.~Jin, H.~Li, and S.-Y. Huang, ``Hpepdock: a web server for blind peptide--protein docking based on a hierarchical algorithm,'' \emph{Nucleic Acids Research}, vol.~46, no.~W1, pp. W443--W450, 2018.

\bibitem{ref24}
J.~L. Watson, D.~Juergens, N.~R. Bennett, B.~L. Trippe, J.~Yim, H.~E. Eisenach, W.~Ahern, A.~J. Borst, R.~J. Ragotte, L.~F. Milles \emph{et~al.}, ``De novo design of protein structure and function with rfdiffusion,'' \emph{Nature}, vol. 620, no. 7976, pp. 1089--1100, 2023.

\bibitem{ref25}
J.~Dauparas, I.~Anishchenko, N.~Bennett, H.~Bai, R.~J. Ragotte, L.~F. Milles, B.~I. Wicky, A.~Courbet, R.~J. de~Haas, N.~Bethel \emph{et~al.}, ``Robust deep learning--based protein sequence design using proteinmpnn,'' \emph{Science}, vol. 378, no. 6615, pp. 49--56, 2022.

\bibitem{ref26}
S.~V{\'a}zquez~Torres, M.~Benard~Valle, S.~P. Mackessy, S.~K. Menzies, N.~R. Casewell, S.~Ahmadi, N.~J. Burlet, E.~Muratspahi{\'c}, I.~Sappington, M.~D. Overath \emph{et~al.}, ``De novo designed proteins neutralize lethal snake venom toxins,'' \emph{Nature}, pp. 1--7, 2025.

\bibitem{ref27}
H.~Liu, Z.~Song, Y.~Zhang, B.~Wu, D.~Chen, Z.~Zhou, H.~Zhang, S.~Li, X.~Feng, J.~Huang \emph{et~al.}, ``De novo design of self-assembling peptides with antimicrobial activity guided by deep learning,'' \emph{Nature Materials}, pp. 1--12, 2025.

\bibitem{ref28}
Q.~Cao, C.~Ge, X.~Wang, P.~J. Harvey, Z.~Zhang, Y.~Ma, X.~Wang, X.~Jia, M.~Mobli, D.~J. Craik \emph{et~al.}, ``Designing antimicrobial peptides using deep learning and molecular dynamic simulations,'' \emph{Briefings in Bioinformatics}, vol.~24, no.~2, p. bbad058, 2023.

\bibitem{ref29}
Y.~Zhao, J.~Yu, Y.~Su, Y.~Shu, E.~Ma, J.~Wang, S.~Jiang, C.~Wei, D.~Li, Z.~Huang \emph{et~al.}, ``A unified deep framework for peptide--major histocompatibility complex--t cell receptor binding prediction,'' \emph{Nature Machine Intelligence}, pp. 1--11, 2025.

\bibitem{ref30}
B.~Dadonaite, J.~Brown, T.~E. McMahon, A.~G. Farrell, M.~D. Figgins, D.~Asarnow, C.~Stewart, J.~Lee, J.~Logue, T.~Bedford \emph{et~al.}, ``Spike deep mutational scanning helps predict success of sars-cov-2 clades,'' \emph{Nature}, vol. 631, no. 8021, pp. 617--626, 2024.

\bibitem{ref31}
N.~Ferruz and B.~H{\"o}cker, ``Controllable protein design with language models,'' \emph{Nature Machine Intelligence}, vol.~4, no.~6, pp. 521--532, 2022.

\bibitem{ref32}
Y.~Gao, Y.~Gao, Y.~Fan, C.~Zhu, Z.~Wei, C.~Zhou, G.~Chuai, Q.~Chen, H.~Zhang, and Q.~Liu, ``Pan-peptide meta learning for t-cell receptor--antigen binding recognition,'' \emph{Nature Machine Intelligence}, vol.~5, no.~3, pp. 236--249, 2023.

\bibitem{ref33}
J.~Huang, Y.~Xu, Y.~Xue, Y.~Huang, X.~Li, X.~Chen, Y.~Xu, D.~Zhang, P.~Zhang, J.~Zhao \emph{et~al.}, ``Identification of potent antimicrobial peptides via a machine-learning pipeline that mines the entire space of peptide sequences,'' \emph{Nature Biomedical Engineering}, vol.~7, no.~6, pp. 797--810, 2023.

\bibitem{ref34}
S.~V{\'a}zquez~Torres, P.~J. Leung, P.~Venkatesh, I.~D. Lutz, F.~Hink, H.-H. Huynh, J.~Becker, A.~H.-W. Yeh, D.~Juergens, N.~R. Bennett \emph{et~al.}, ``De novo design of high-affinity binders of bioactive helical peptides,'' \emph{Nature}, vol. 626, no. 7998, pp. 435--442, 2024.

\bibitem{ref35}
Y.~Lei, S.~Li, Z.~Liu, F.~Wan, T.~Tian, S.~Li, D.~Zhao, and J.~Zeng, ``A deep-learning framework for multi-level peptide--protein interaction prediction,'' \emph{Nature Communications}, vol.~12, no.~1, p. 5465, 2021.

\bibitem{ref36}
Q.~Deng, Z.~Wang, S.~Xiang, Q.~Wang, Y.~Liu, T.~Hou, and H.~Sun, ``Rlpmiec: High-affinity peptide generation targeting major histocompatibility complex-i guided and interpreted by interaction spectrum-navigated reinforcement learning,'' \emph{Journal of Chemical Information and Modeling}, vol.~64, no.~16, pp. 6432--6449, 2024.

\bibitem{ref37}
S.~Chen, T.~Lin, R.~Basu, J.~Ritchey, S.~Wang, Y.~Luo, X.~Li, D.~Pei, L.~B. Kara, and X.~Cheng, ``Design of target specific peptide inhibitors using generative deep learning and molecular dynamics simulations,'' \emph{Nature Communications}, vol.~15, no.~1, p. 1611, 2024.

\bibitem{ref38}
J.~Ho, A.~Jain, and P.~Abbeel, ``Denoising diffusion probabilistic models,'' \emph{Advances in Neural Information Processing Systems}, vol.~33, pp. 6840--6851, 2020.

\bibitem{ref39}
R.~Yu, D.~J. Craik, and Q.~Kaas, ``Blockade of neuronal $\alpha$7-nachr by $\alpha$-conotoxin imi explained by computational scanning and energy calculations,'' \emph{PLoS Computational Biology}, vol.~7, no.~3, p. e1002011, 2011.

\bibitem{ref40}
A.~Elnaggar, M.~Heinzinger, C.~Dallago, G.~Rehawi, Y.~Wang, L.~Jones, T.~Gibbs, T.~Feher, C.~Angerer, M.~Steinegger \emph{et~al.}, ``Prottrans: towards cracking the language of life’s code through self-supervised learning,'' \emph{IEEE Transactions on Pattern Analysis and Machine Intelligence}, vol.~44, pp. 7112--7127, 2021.

\bibitem{ref41}
X.~Wu, A.~J. Hone, Y.-H. Huang, R.~J. Clark, J.~M. McIntosh, Q.~Kaas, and D.~J. Craik, ``Computational design of $\alpha$-conotoxins to target specific nicotinic acetylcholine receptor subtypes,'' \emph{Chemistry--A European Journal}, vol.~30, no.~7, p. e202302909, 2024.

\bibitem{ref42}
R.~Wu, F.~Ding, R.~Wang, R.~Shen, X.~Zhang, S.~Luo, C.~Su, Z.~Wu, Q.~Xie, B.~Berger \emph{et~al.}, ``High-resolution de novo structure prediction from primary sequence,'' \emph{Preprint at https://doi.org/10.1101/2022.07.21.500999, BioRxiv}, 2022.

\bibitem{ref43}
J.~Jumper, R.~Evans, A.~Pritzel, T.~Green, M.~Figurnov, O.~Ronneberger, K.~Tunyasuvunakool, R.~Bates, A.~{\v{Z}}{\'\i}dek, A.~Potapenko \emph{et~al.}, ``Highly accurate protein structure prediction with alphafold,'' \emph{Nature}, vol. 596, no. 7873, pp. 583--589, 2021.

\bibitem{ref44}
J.~Xu and Y.~Zhang, ``How significant is a protein structure similarity with tm-score= 0.5?'' \emph{Bioinformatics}, vol.~26, no.~7, pp. 889--895, 2010.

\bibitem{ref45}
J.~Kyte and R.~F. Doolittle, ``A simple method for displaying the hydropathic character of a protein,'' \emph{Journal of Molecular Biology}, vol. 157, no.~1, pp. 105--132, 1982.

\bibitem{ref46}
Z.~Lin, H.~Akin, R.~Rao, B.~Hie, Z.~Zhu, W.~Lu, N.~Smetanin, R.~Verkuil, O.~Kabeli, Y.~Shmueli \emph{et~al.}, ``Evolutionary-scale prediction of atomic-level protein structure with a language model,'' \emph{Science}, vol. 379, no. 6637, pp. 1123--1130, 2023.

\bibitem{ref47}
L.~McInnes, J.~Healy, and J.~Melville, ``Umap: Uniform manifold approximation and projection for dimension reduction,'' \emph{arXiv preprint arXiv:1802.03426}, 2018.

\bibitem{ref48}
N.~Tabassum, H.-S. Tae, X.~Jia, Q.~Kaas, T.~Jiang, D.~J. Adams, and R.~Yu, ``Role of cysi--cysiii disulfide bond on the structure and activity of $\alpha$-conotoxins at human neuronal nicotinic acetylcholine receptors,'' \emph{ACS Omega}, vol.~2, no.~8, pp. 4621--4631, 2017.

\bibitem{ref49}
D.~A. Case, H.~M. Aktulga, K.~Belfon, D.~S. Cerutti, G.~A. Cisneros, V.~W.~D. Cruzeiro, N.~Forouzesh, T.~J. Giese, A.~W. G{"o}tz, H.~Gohlke \emph{et~al.}, ``Ambertools,'' \emph{Journal of Chemical Information and Modeling}, vol.~63, no.~20, pp. 6183--6191, 2023.

\end{thebibliography}

\end{document}